\documentclass[letter]{aa} 
\usepackage[figuresright]{rotating}
\usepackage[authoryear]{natbib}
\bibpunct{(}{)}{;}{a}{}{,}

\usepackage{graphicx}
\usepackage{txfonts}
\newcommand{\flux}{erg cm$^{-2}$ s$^{-1}$}

\begin{document}
   \title{A  candidate tidal disruption  event in the Galaxy cluster Abell 3571}

   \author{N. Cappelluti
          \inst{1,2}
          \and
	M. Ajello\inst{3,4}
	\and
	P. Rebusco\inst{5}
	\and
          S. Komossa\inst{1}
	\and
	A. Bongiorno\inst{1}
	\and
	C. Clemens\inst{1}
	\and
	M. Salvato\inst{6}
	\and
	P. Esquej\inst{7,1}
	\and
	T. Aldcroft\inst{8}
	\and
	J. Greiner\inst{1}
	\and
	H. Quintana\inst{9}		
	          }

   \offprints{N. Cappelluti}
	 \institute{ Max-Planck-Institute f\"ur Extraterrestrische Physik, Postfach 1312,
               85741, Garching bei M\"unchen, Germany
              \email{cap@mpe.mpg.de}
         \and
	 University of Maryland, Baltimore County, 
	 1000 Hilltop Circle, Baltimore, MD 21250.
	\and
	SLAC National Accelerator Laboratory, 2575 Sand Hill Road, Menlo Park, CA 94025, USA 
	\and
	 KIPAC , 2575 Sand Hill Road, Menlo Park, CA 94025, USA     
	\and
	Kavli Institute for Astrophysics and Space Research, MIT, Cambridge, MA 02139, USA 
	\and
	California Institute of Technology, 105-24 Robinson, 
	1200 East California Boulevard, Pasadena, CA 91125.
	\and
	Dept. of Physics and Astronomy, Leicester University, Leicester LE1 7RH, U.K.
	\and
	  Harvard-Smithsonian Center for Astrophysics, 60 Garden St, Cambridge, MA 02138
        \and
       	 Department of Astronomy and Astrophysics, Pontificia Universidad Catolica de Chile. Casilla 306, Santiago 22, Santiago, Chile. 
	}

   \date{}

 
  \abstract
   {Tidal disruption events are possible sources of temporary  nuclear activity 
	in  galactic nuclei and can be considered as  good indicators of the existence of 
	supermassive black holes in  the centers of  galaxies.     }  
   {A new X-ray source has been detected serendipitously with ROSAT  in a PSPC pointed observation of the 
	galaxy cluster A3571. Given the strong flux decay of the object in subsequent detections, the tidal disruption scenario
	is investigated as a possible explanation of the event.}
   {We followed the evolution of the X-ray transient with 
    ROSAT, XMM-Newton and Chandra for a total period of $\sim$13 years. We also obtained 7-band optical/NIR photometry 
	with GROND at the ESO/MPI 2.2m telescope.}
   {We report a very large decay of the X-ray flux of the ROSAT source  identified with  the galaxy LEDA 095953, a member of 
    the cluster Abell 3571. We measured a  maximum 0.3--2.4 keV luminosity Log(L$_{X}$)=42.8 erg s$^{-1}$. The high state of the 
	source lasted at least 150 ks; afterwards  L$_{X}$ declined as$\sim$ t$^{-2}$. 
	The spectrum of the brightest epoch  is consistent with a black body with temperature $kT\sim$0.12 keV. }
         { The total energy released  by this event in 10 yr was estimated to be $\Delta~E>$2$\times$10$^{50}$ erg.	 
	 We interpret this event as a tidal disruption of a solar type star by the central
	 supermassive black hole (i.e. $\sim$10$^{7}$ M$_{\odot}$)
	 of  the galaxy.    }
   {}

   \keywords{ X-rays -- Galaxies, X-rays:  bursts, Galaxies: clusters: individual: A3571, Galaxies: Active, Galaxies: Nuclei 
                --
               }
\maketitle

\section{Introduction}
There is  growing evidence that most  galaxies
with a bulge host a supermassive black hole (BH)  in their center \citep{Kor,mago}. 
The most recent theoretical and observational developments in the field  
of galaxy evolution suggest that during their life, galaxies experience one or more phases 
where the BH is active and  powered by accretion. One of the effects  
of this phenomenon  are the  so called active galactic nuclei or quasars where accretion
occurs on time scales longer than 10$^{6-7}$yr. 
It is widely accepted that during this period the BH is fed by an  accretion disk.
On the other hand,  during the non-active phase of the galaxy life the BHs
can be briefly  powered  by stars tidally disrupted during
close encounters with the BHs \citep[e.g.][]{frank}.  Recent studies 
\citep[see e.g.][]{wang,esquej,donley} pointed out that 
 these events  happen with a rate of
$\sim$10$^{-4}$-10$^{-5}$ galaxy$^{-1}$ year$^{-1}$. 
These phenomena can  generate strong flares 
in the UV-X-ray band as the result of the 
onset of  an accretion disk.
The X-ray/UV  luminosity of the galaxy 
is therefore expected to grow to values
close to those of  AGNs (i.e. $>$10$^{42}$ erg s$^{-1}$).
If the   star is  Sun-like, tidal disruption  can  only occur 
if the mass of the BHs does not exceed $\sim$10$^{8}$ M$_{\odot}$ since, 
in this case the tidal radius would typically fall within the Schwarzschild radius
and the star is swallowed as a whole. In the case of rotating BHs, the tidal disruption 
also can occur for higher BH masses if the star approaches from a favorable 
direction \citep{belo}. However atmosphere stripping of 
giant stars is  possible  also for larger BH masses.
 When assuming a black body radiating
at the Eddington luminosity, the temperature of the emitting
region calculated at the last stable orbit (3 R$_{S}$, where R$_{S}$ is
the Schwarzschild radius) is $\sim$30-300 eV for a 10$^{6-7}$ M$_{\odot}$ black
hole.  After the disruption, the mass accretion rate, and therefore  the bolometric luminosity,
may decay approximately  as t$^{-5/3}$ \citep{rees,evans}, even though deviations from this law
are expected \citep[see e.g.][]{sempresia}. 
7 candidate tidal disruption events have been  detected 
in the X-ray by the
{\em ROSAT} all-sky survey and in the XMM-Newton slew survey 
\citep[see e.g][]{komossa99,greiner,grupe,komossa04,esquej07}. 
Candidate tidal disruption have been observed also in the Optical/UV 
 \citep[e.g.][]{renzini,gezari}. 
The X-ray events show  strong X-ray variability (i.e. more than a factor of 100),
0.3--2.4 keV luminosities typical of AGN and quasars  
(i.e. $>$10$^{42}$ erg s$^{-1}$) declining as $\sim$t$^{-5/3}$, ultrasoft (i.e. $\sim$0.1 keV)  X-ray 
spectra and the absence of Seyfert activity in ground-based optical
spectra (see Komossa 2002 for a review). 
In this letter we report the serendipitous detection of a very large decay in the
X-ray flux  of the galaxy  LEDA 095953 at z=0.0366 \citep{quintana}
in the field of view of the galaxy cluster Abell 3571 using {\em ROSAT}, XMM-Newton 
and Chandra archival X-ray  data and GROND optical/NIR follow-up. 
\begin{figure}[!t]
\begin{center}
\resizebox{\hsize}{!}
{\includegraphics{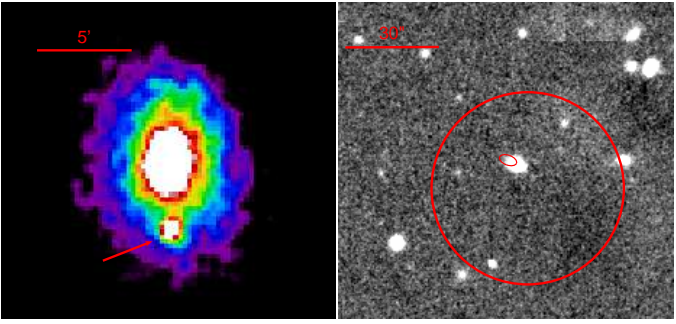}}
\caption{\label{fig:images} $ Left~Panel$: The {\em ROSAT}-PSPC 0.3-2.4 keV colour 
coded image of the field of A3571. 
The X-ray transient is indicated with the arrow. $Right~Panel$: 
 GROND image in the K$_{S}$-band of the 
the region of the X-ray transient, the $red~circle$ represents the 
  {\em ROSAT}-PSPC error-box and the $red~ellipsoid$ is the Chandra confidence region.}
\end{center}
\end{figure}
\section{Observations and data processing}
\begin{table}[!b]
\tiny
\begin{tabular}{cccc}
\hline
\hline
Instrument &  Date & Band & Exposure \\
           & dd-mm-yyyy    &       & s \\ 
\hline
ROSAT-PSPC & 12-08-1992	   & 0.3--2.4 keV      &   5800\\
ROSAT-HRI  & 05-08-1994    & 0.3--2.4 keV      &  19200\\
XMM-{\em Newton}-EPIC-PN & 29-07-2002 & 0.3--2.4 keV       & 24200\\
{\em Chandra}-ACIS-S & 31-07-2003 & 0.3--2.4 keV          & 33900\\
ESO/MPI-GROND & 12-08-2008   &g, r',i', z' & 458  \\
ESO/MPI-GROND & 12-08-2008   & J,H,K & 480  \\
\hline
\end{tabular}
\caption{\label{tab:log} The log of the X-ray and optical/NIR observation of the  field of Abell 3571 used for this work. }
 \end{table}
During the soft X-ray follow-up of the   Swift-BAT, hard X-ray selected  sample of galaxy
clusters \citep{ajello}, we serendipitously discovered a bright source 
in a {\em ROSAT}-PSPC image of Abell 3571 (see Fig. \ref{fig:images})
that apparently disappeared in subsequent observations with  XMM-{\em Newton}  and {\em Chandra}. 
The field of Abell 3571 has been observed in  X-rays in pointed mode with the {\em ROSAT}-PSPC/HRI 
detectors, XMM-{\em Newton} EPIC, {\em Chandra} ACIS-S and during 
the   {\em ROSAT} all-sky survey. 
The {\em ROSAT} data were processed using the standard MIDAS-EXSAS software version 03SEPpl1, 
while XMM-Newton and Chandra data were processed using XMM-SAS version 7.0 and CIAO 4.0.
We also performed an optical/NIR follow-up  of the field in the 
g', r, i', z' bands and J, H, K$_{S}$ bands, by using  GROND  
\citep{grond}, mounted at the 2.2 m ESO/MPI telescope 
at La Silla observatory (Chile).  In Table \ref{tab:log} we list the  log of the observations
of the field of Abell 3571 relevant for this work.
\subsection{Data analysis}
\begin{figure}[!t]
\begin{center}
\resizebox{0.7\hsize}{!}
{\includegraphics[angle=270]{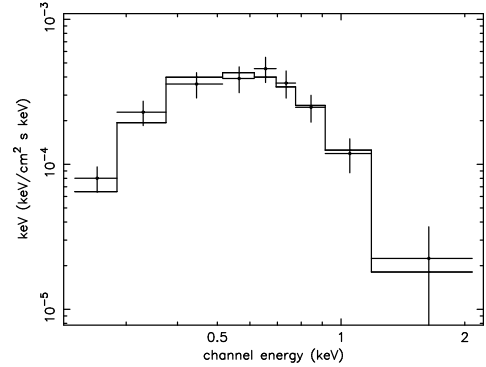}}
\caption{\label{fig:spectrum} $Left~Panel$: 
The {\em ROSAT}-PSPC unfolded spectrum of the source TDXFJ134730.3-325451 ($Crosses$) and the 
best fit blackbody model $Continuous~Line$.   }
\end{center}
\end{figure}

\begin{figure}[!t]
\begin{center}
\resizebox{0.65\hsize}{!}
{\includegraphics{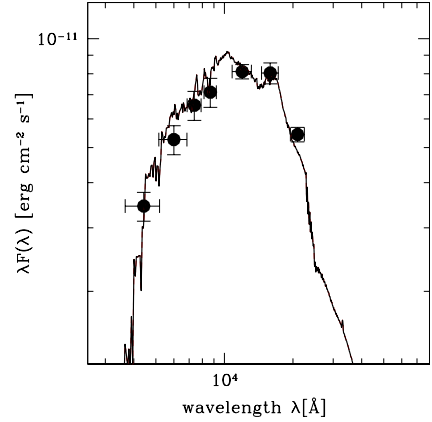}}
\caption{\label{fig:sed} The  SED of the galaxy LEDA 095953 in the g', r', i', z', H, J and K$_{S}$
bands obtained with GROND over the best fit SED  model of a S0 galaxy ($solid~line$).}
\end{center}
\end{figure}

{\bf {\em ROSAT}-PSPC}: Using the {\em ROSAT}-PSCP  0.3-2.4 keV energy band 
image and exposure map available in the archive, we created a background map,
 by first  running a sliding cell source detection 
  with   a signal-to-noise ratio threshold 
of  4, excising the detected sources from the image and
fittingthe residual data with a spline model.
With the background map described above and using 
the exposure map we run a 
maximun likelihood, PSF-fitting,  source detection. 
The bright, eye-detected, source mentioned in the previous section 
has been detected with a significance $>$10$\sigma$
 with coordinates of the centroid 13$^{h}$47$^{m}$29.8$^{s}$ -32$^{\circ}$55$^{\arcmin}$00$^{\arcsec}$ 
(J2000) with an error box of $\sim$25$\arcsec$.
Considering that the source is embedded in the X-ray emission 
of the galaxy cluster Abell 3571, the photometric properties
have been evaluated using aperture photometry instead of using the estimate of the
detection software.
 We extracted the source counts in a circular region with a radius corresponding to
the 90\% Encircled Energy Radius (EER)  around the centroid estimated by the maximum likelyhood fit.
Background counts have been estimated in a ring around the source
with an inner radius 1.5 times the 90\% EER and outer radius 2 times the 90\% ERR.
As a result we obtained 305$\pm{27}$ net source counts corresponding  to a count-rate of (5.2$\pm{0.4}$)$\times$10$^{-2}$ s$^{-1}$. 
PSF fitting estimated a flux $\sim$30\% lower as a result of a likely overestimate of the background.
In the same regions we also extracted  source and background  lightcurves and spectra. 
All the epochs but one have flux  consistent with a mean  count-rate of $\sim$5.2$\times$10$^{-2}$ s$^{-1}$.
Note that the observation was split in two parts separated by $\sim$150 ks. This indicates 
that no strong variability is present  on such a time scale.
However in one time bin  the count-rate  reached 
(8.3$\pm{1.0}$)$\times$10$^{-2}$ s$^{-1}$. This flux is $\sim$3$\sigma$ above the mean
and could indicate a small flare  during the event.  \\ 
The unfolded spectrum of the source  is shown in Fig. \ref{fig:spectrum} . 
We performed a $\chi^{2}$ fit to
the spectrum using a black-body model plus cold absorption.  
We kept as free parameters the temperature kT, the columns density
N$_{H}$ and the normalization $k$. 
As a result we obtained a temperature kT=120$\pm{16}$ eV and
  N$_{H}$=3.93$_{-1.24}^{+1.43}\times$10$^{20}$ cm$^{-2}$ and $\chi^{2}/\nu$=1.05/4.00.
 This corresponds to a  0.3-2.4 keV flux of 7.65$\pm{0.92}\times$10$^{-13}$ \flux.
 Assuming no spectral changes during the observation, the maximum
0.3-2.4 keV flux emitted by the source is 1.16$\pm{0.13}\times$10$^{-12}$ \flux. 
Note that the column density is consistent within 
$1\sigma$ with  the Galactic value predicted by \citet{lock82} 
(i.e. N$_{H}$=3.7$\times$10$^{20}$ cm$^{-2}$) suggesting an extragalactic origin of the source.
We also tested other spectral models, like  a power-law plus absorption, without improving the
quality of the fit by having $\Gamma\sim$6.5, and
 N$_{H}\sim$3$\times$10$^{21}$ cm$^{-2}$ as best fit parameters and $\chi^{2}/\nu$=5.30/4.00.\\
We also performed aperture photometry on {\em ROSAT} all sky survey images 
at the source location and we determined a 1$\sigma$ upper-limit (see next Section)
on the count-rate of 0.44 s$^{-1}$ corresponding to
 a flux of $\sim$2.2$\times$10$^{-12}$ \flux~in the 0.3--2.4 keV band
assuming as a  spectral model a black body with kT=0.12 keV.\\
{\bf {\em ROSAT}-HRI}: The source is visually observable also in the HRI observation
though much fainter than in the PSPC image. Considering the highest spatial resolution of the HRI detector we performed
a source detection to constrain the source position better than in the PSPC pointing. 
Unfortunately because of the faintness of the source and the difficulty in modeling 
the high background due to the extended X-ray emission of 
the cluster, the source has been detected with a significance of $\sim$3-4$\sigma$, but 
as an extended source. We therefore performed aperture photometry around the {\em ROSAT}-PSPC centroid.
By adopting the same procedure used for the PSPC data,
we measured  38$\pm{9}$ source counts corresponding to a count rate of 
(1.97$\pm{0.45}$)$\times10^{-3}$ s$^{-1}$.
By assuming no spectral change between the {\em ROSAT}-PSPC and 
HRI observation (i.e. 2 years) we estimated a 0.3--2.4 keV  flux of
1.53$\pm{0.4}\times$10$^{-13}$ \flux.\\
{\bf XMM-Newton:} A visual inspection of the EPIC-PN 
image does not show any relevant signature of a X-ray 
source in the {\em ROSAT}  error box. We run a maximum likelihood 
source detection in that region but no sources were detected
with significance $>$3$\sigma$. 
We therefore estimated the 3$\sigma$ upper-limit of the source count at the  source location by
using the prescriptions of \citet{nar06}.
Given M counts actually observed in a region of 30$\arcsec$ \citep{cappelluti},
and B background counts, the 1$\sigma$ 
upper limit is defined as the number of counts $X$ that gives the probability
to observe M (or fewer) counts equal to the formal 68.3\%  Gaussian probability:
$P(\leq M, X+B)=P_{Gauss}$(68.3\%).Assuming Poissonian statistics this equation becomes:
$P_{Gauss}=e^{-(X+B)}\sum_{i=0}^{M}\frac{(X+B)^{i}}{i!}.$
By  iteratively solving the previous equation  in the case of $P_{Gauss}$=0.997,
we obtained the 3$\sigma$ upper limit X=167 counts.
This corresponds to a count-rate of 8.8$\times$10$^{-3}$ s$^{-1}$ and to a flux of 1.45$\times$10$^{-14}$\flux\footnote{
The flux was derived from the count-rate with the PIMMS software by assuming as the spectrum the 
{\em ROSAT}-PSPC best fit.  {\bf http://heasarc.nasa.gov/Tools/w3pimms.html}  }.\\
{\bf Chandra:} Using the task {\em wavdetect},
included in the CIAO software, we performed a
source detection on the ACIS-S4 Chip where the source is observable. 
The analysis was performed in the 0.3--2.4 keV energy band. 
We ran  {\em wavdetect} using an exposure map estimated at 1 keV with a threshold of 10$^{-6}$, corresponding to a maximum
 of 1 spurious detection expected per ACIS-chip. 
As a result, in the {\em ROSAT} PSPC error box we detected one source with 9.5$\pm{3.74}$ counts
corresponding to a count-rate of  2.89$\times$10$^{-4}$ s$^{-1}$. The new source coordinates
are $\alpha$=13$^{h}$47$^{m}$30.33$^{s}$, $\delta$=-32$^{\circ}$54${\arcmin}$50.63${\arcsec}$. 
At the source position the estimated $\sigma$ positional error is $\sim$~2.5$\arcsec$.
We also ran a source detection in the 2--7 keV band without detecting a source. 
We also estimated the flux of the source by folding into XSPEC the Chandra response matrices 
and assuming as the spectral model the {\em ROSAT}-PSPC best fit and obtained a 0.3--2.4 keV flux of 
3.43$\pm{1.35}\times$10$^{-15}$ \flux. 
Note that the source centroid lies about 5$\arcsec$ from the chip edge, therefore
the estimate of the background could be affected by features near this edge.
According to its coordinate, the X-ray source has been named TDXFJ134730.3-325451. \\
{\bf GROND:}
In Fig. \ref{fig:images} we show the  GROND K-band image
of the area covered by the  {\em ROSAT}-PSPC error box. 
As one can notice, the Chandra ellipsoid
is placed right on the disk of a bright galaxy
with an offset of $\sim$3 \arcsec~ from the optical
centroid of the galaxy. Note that the GROND images have an 
astrometric uncertainty of 0.5\arcsec.
We searched the NED catalog and the galaxy  
was previously known  as   LEDA 095953, a  member of the cluster Abell 3571
at z=0.0366 \citep{quintana}.  
 We determined the photometric properties of the source 
 and  obtained the following AB magnitudes: {\em 17.05$\pm{0.10}$, 16.26$\pm{0.10}$,
15.80$\pm{0.10}$, 15.54$\pm{0.10}$, 15.04$\pm{0.02}$, 14.75$\pm{0.07}$ and 14.87$\pm{0.04}$} in 
the  g' ,r', i',  z', J, H, K$_{S}$ bands, respectively. These magnitudes are calculated by using
SA105-815 and 2MASS  stars as a reference and corrected for the expected foreground 
extinction of $E_{B-V}$=0.054 \citep{est}.
We examined the global spectral energy distribution (SED) in Fig. \ref{fig:sed}
of this object using a tool based on a chi-squared minimization method 
that allows us to fit the observed fluxes with a combination of AGN and galaxy emission,
 also allowing  for the possibility of extinction of the AGN flux. 
From this analysis we found that the model fit that better describes the 
SED of this object is a normal S0 galaxy with no significant 
contribution due to the presence of an AGN.
\section{Long term light curve}
\begin{figure}[!t]
\begin{center}
\resizebox{0.65\hsize}{!}
{\includegraphics{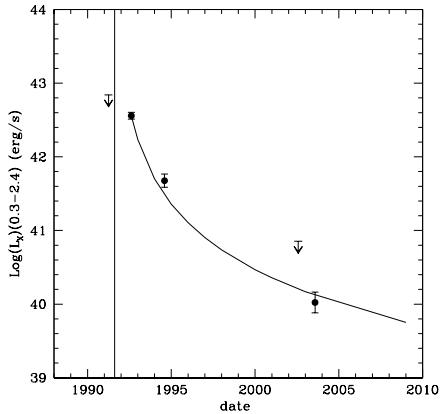}}
\caption{\label{fig:ltlc} The X-ray lightcurve of  TDXFJ134730.3-325451. 
The $solid~curve$ represents the result of the power-law fit
while the $vertical~line$ marks the expected begin of the flare. }
\end{center}
\end{figure}

The X-ray flux  of TDXFJ134730.3-325451 has been followed for 13 years and
	its lightcurve is presented in Fig. \ref{fig:ltlc}. Remarkably, it
shows an amplitude of the variability of a factor $>$650.
We estimated the maximum 0.3-2.4 keV luminosity\footnote{The luminosity has been computed 
assuming a $\Lambda$-dominated universe with $\Omega_{\Lambda}$=0.7, $\Omega_{m}$=0.3 and
H$_{0}$=72 km s$^{-1}$ Mpc$^{-1}$} 
as Log(L$_X$)=42.83 erg s$^{-1}$.
By excluding the upper limits, we fitted the long term lightcurve with a power-law
 model in the form:
  $L_{X}(t)=k*(\frac{t-t_{disr}}{1 yr})^{-\alpha}$,
where k is the normalization, $t_{disr}$ marks the time of the star's disruption 
and $\alpha$ is the slope of the decay. 
As a result we  obtained k$\sim$7$\times$10$^{42}$ erg s$^{-1}$,
$t_{disr}$$\sim$1991.6 and $\alpha$=2.2$\pm{0.5}$. 
Because of the low number of data points, the uncertainties
on the fit parameters are large.  The value of the slope $\alpha$ is therefore
marginally consistent with the predictions of variants of tidal 
disruption models \citep[i.e. $\alpha\sim$5/3, see e.g.][]{evans,rees}. 
Note that according to the lightcurve, the source at the maximum could have been
at  1-2 orders of magnitude brighter with respect to the
time of the {\em ROSAT}-PSPC observation.
 If the source continues  its decay, it would be detectable with Chandra for all the year 2009 and 2010,
with moderately deep exposures.
\section{Discussion}
We serendipitously detected a factor $>$650  decay in the
flux of an X-ray source in the rich galaxy cluster A3571. 
The flare  most likely originated in 
the galaxy LEDA 095953 at z=0.0366. The  most luminous epoch in the 
lightcurve reached  Log${L_{X}}$=42.83 (0.3--2.4) and lasted
for more than 150 ks, although higher luminosities  at the time of the burst cannot
be excluded. 
The spectrum of  the flare  is consistent with a  black-body of temperature 
kT$\sim$0.12 keV with no significant hard components.
The optical/NIR SED of the host galaxy does not show any  signature 
of AGN activity.  In addition to this, the decay of the long term light
curve is marginally consistent with t$^{-5/3}$, 
and thanks to the stringent RASS upper limit, 
we likely  detected the source   close to the burst.
The X-ray luminosity and the variability of the source 
are therefore much higher than  those of
ULX, which have  L$_{X}<$10$^{39-40}$ erg s$^{-1}$ and  flux variations
 typically of up to a factor of 2-3 \citep[see e.g.][]{ulx}.
Supernovae  explosions can also produce X-ray flares,
but these are either of very short duration (hours), or else if longer lasting
they have much harder X-ray spectra and typically have
L$_{X}$$<$10$^{41}$ erg s$^{-1}$\citep[see e.g.][for a review]{immler}.
Additionally novae and supersoft sources can produce  bursts but with much shorter duration \citep{tanaka}.
According to  the event properties, we can  safely exclude these kinds of 
stellar phenomena.
However, such a strong variability could be observed in GRB afterglows 
or Hypernovae but with a much shorter duration of the outburst.
The characteristics of the flare are therefore consistent 
with previous cases of candidate tidal disruption events \citep[see e.g.][]{komossa04,esquej}
and with theoretical predictions \citep{rees,sempresia}. 
The mass of the BH should be not much higher than $\sim$10$^{8}$ M$_{\odot}$
to disrupt a Sun-like star outside the Schwarzschild radius and to allow the onset 
of accretion. 
We  then estimated  the black hole mass from the L$_{K}$-M$_{BH}$ relation determined 
by \citet{marconi} in the local Universe.  If L$_{K}$ is the K-band luminosity of the bulge of the galaxy
in solar units then $\log(M_{BH}/M_{\odot})$$\sim$8.08+1.21( L$_{K_{bulge}}$/L$_{\odot}$-10.9).
Using an aperture radius of $\sim$4$\arcsec$ (i.e. $\sim$2.5 kpc, at the galaxy distance)
we estimated a log(L$_{K_{bulge}})$$\sim$10$L/L_{\odot}$ which corresponds to
 $\log(M_{BH}/M_{\odot})$$\sim$7.
The total energy released during  the event can be estimated with $\Delta E_{X}$=$\int_{t}^{\infty}L_{X}(t)dt$,
where $L_{X}$(t) is the best fit to the light curve.
  By integrating over a period of $\sim$10 yr after our first data point   we obtained a total energy 
release of the order of 2$\times$10$^{50}$ erg. If $\epsilon$$\sim$0.1 is 
the typical efficiency of mass to energy conversion for accretion onto  a BH,
then the total mass accreted during a 10 year period 
is $\Delta M$$\sim \Delta E_{X}/\epsilon c^{2}$. This yields 
a total mass deposition of 1$\times$10$^{-3}$ M$_{\odot}$.
This value is actually a lower limit, since we started
the integration only at the start of the  observations 
(note that the fitting function diverges at the time of the disruption).
Moreover it is possible   that the total emission 
is dominated by an unobserved EUV component (extrapolation 
of the single black body fit itself predicts a 7\% higher luminosity), and
the true accretion rate is much higher.
According to the results of \citet{li}, for a black hole of
10$^{7}$ M$_{\odot}$,  the flare occurs $\sim$0.07 yr after the disruption.
A 10 yr integration of the lightcurve model from this epoch yields an estimate of the
 accreted mass of  $\sim$0.03 M$_{\odot}$. 
A numerical simulation by \citet{ayal} showed that for a 1 M$_{\odot}$ star disrupted
by a 10$^{6}$ M$_{\odot}$ BH, only a fraction of the order of $\sim$10\% of the initial mass
is actually accreted. Our  limits  on the accreated mass are therefore consistent
with the scenario of a disruption of a 1 M$_{\odot}$ star by the BH.
However a  partial stellar disruption or an explosive disruption \citep{brassart} cannot be ruled out, although 
 disruption and accretion of a brown dwarf or of
  a planets  is very unlikely. We note that since the tidal 
disruption rates are likely of the order of
10$^{-4}$-10$^{-5}$ galaxy$^{-1}$ yr$^{-1}$,   galaxy clusters
are ideal laboratories to explore this phenomenon:
by conservatively assuming 100 member galaxies per cluster \citep{reip}, 
one should expect 1 tidal disruption event per cluster every 
10-100 yrs.  The eROSITA all-sky survey \citep{rosita}  will detect 
several thousands  of galaxy clusters and therefore a large number of these
events.
\acknowledgements
The XMM-Newton, Chandra and ROSAT archival systems and the full 
GROND team are acknowledged. PR is supported by the Pappalardo fellowship at MIT.
HQ thanks partial support from the FONDAP Centro de Astrofisica.

\end{document}